\documentclass[12pt]{article}

\usepackage{sgame}
\usepackage{color}

\usepackage{changebar}
\usepackage{url}
\usepackage{amsmath}
\usepackage{amssymb}
\usepackage{handbookbib}

\usepackage{color}
\usepackage{graphics}

\newcommand{\GP}{\textit{GP}}
\newcommand{\LP}{\textit{LP}}
\newcommand{\GU}{\textit{GU}}
\newcommand{\LU}{\textit{LU}}
\newcommand{\GC}{\textit{GC}}
\newcommand{\LC}{\textit{LC}}

\newcommand{\GR}{\textit{GR}}
\newcommand{\LR}{\textit{LR}}
\newcommand{\GS}{\textit{GS}}
\newcommand{\LS}{\textit{LS}}
\newcommand{\MGS}{\textit{MGS}}
\newcommand{\MLS}{\textit{MLS}}
\newcommand{\GW}{\textit{GW}}
\newcommand{\LW}{\textit{LW}}
\newcommand{\GF}{\textit{GF}}
\newcommand{\LF}{\textit{LF}}
\newcommand{\MGW}{\textit{MGW}}
\newcommand{\MLW}{\textit{MLW}}

\newcommand{\oldbfe}[1]{\begin{bfseries}\emph{#1}\end{bfseries}}

\newcommand{\ES}{\mbox{$\emptyset$}}

\newcommand{\myra}{\mbox{$\:\rightarrow\:$}}

\newcommand{\A}{\mbox{$\ \wedge\ $}}

\newcommand{\sse}{\mbox{$\:\subseteq\:$}}

\newcommand{\fa}{\mbox{$\forall$}}
\newcommand{\te}{\mbox{$\exists$}}

\newcommand{\LL}{\mbox{$\ldots$}}

\newcommand{\C}[1]{\mbox{$\{{#1}\}$}}           

\newcommand{\NI}{\noindent}
\newcommand{\HB}{\hfill{$\Box$}}
\newcommand{\VV}{\vspace{5 mm}}

\newcommand{\II}{\vspace{2 mm}}

\newcommand{\szkew}[1]{\relax \setbox0=\hbox{\kern -24pt $\displaystyle#1$\kern 0pt }%
\box0}
{\catcode`\@=11 \global\let\ifjusthvtest@=\iffalse}

\newcounter{oldmycaption}

\def\smallromani{\renewcommand{\theenumi}{\roman{enumi}}
\renewcommand{\labelenumi}{(\theenumi)}}

\newcommand{\Proof}{\NI
                    {\bf Proof.}\ }
\newtheorem{theorem}{Theorem}
\newtheorem{defined}{Definition}

\newtheorem{exa}{Example}
\newenvironment{example}{\begin{exa} \rm}{\end{exa}}

\newtheorem{lemma}{Lemma}
\newtheorem{corollary}{Corollary}

\usepackage{amsfonts}
\usepackage{latexsym}
\usepackage{alltt}

\title{Relative Strength of Strategy Elimination Procedures}

\author{Krzysztof R. Apt \\
\emph{CWI, Amsterdam, the Netherlands} \\
and \emph{University of Amsterdam}
}

\date{}
\begin{document}

\maketitle

\begin{abstract}
  We compare here the relative strength of four widely used procedures
  on finite strategic games: iterated elimination of weakly/strictly
  dominated strategies by a pure/mixed strategy.  A complication is
  that none of these procedures is based on a monotonic operator.  To
  deal with this problem we use 'global' versions of these operators.
  
  
\end{abstract}

\section{Introduction}

In the literature four procedures of reducing finite strategic games
have been widely studied: iterated elimination of weakly/strictly
dominated strategies by a pure/mixed strategy.  Denote the
corresponding operators (the mnemonics should be clear, '$L$' refers
to 'local' the meaning of which will be clarified later) respectively
by $\LW$, $\MLW$, $\LS$ and $\MLS$.  When these operators are applied
to a specific game $G$ we get the following obvious inclusions:
\[
\mbox{$\MLW(G) \sse \LW(G) \sse \LS(G)$ and $\MLW(G) \sse \MLS(G) \sse \LS(G)$.}
\]

It is then natural to expect that these inclusions carry on to the
outcomes of the iterations of these operators. It turns out
that this is not completely true. Moreover, proofs of some of the
apparently obvious implications are not, in our view, completely
straightforward.  One of the complications is that none of these
operators is monotonic.  To reason about them we use their
`global' versions.

More precisely, given two strategy elimination operators $\Phi_l$ and
$\Psi_l$ such that for all games $G$, $\Phi_l(G) \sse \Psi_l(G)$ we
prove the inclusion $\Phi^{\omega}_l \sse \Psi^{\omega}_l$ between the
outcomes of their iterations by means of the following generic
procedure:

\begin{enumerate} \smallromani
\item define the corresponding `global' versions of these operators,
$\Phi_g$ and $\Psi_g$,

\item prove that $\Phi^{\omega}_g = \Phi^{\omega}_l$ and $\Psi^{\omega}_g = \Psi^{\omega}_l$,

\item show that for all games $G$, $\Phi_g(G) \sse \Psi_g(G)$,

\item show that at least one of $\Phi_g$ and $\Psi_g$ is monotonic.
\end{enumerate}

The last two steps then imply $\Phi^{\omega}_g \sse \Psi^{\omega}_g$
by a general lemma. The desired inclusion $\Phi^{\omega}_l \sse \Psi^{\omega}_l$ then
follows by (ii). The main work is in proving (ii). 
%
\section{Preliminaries}

\subsection{Strategic games}

By a \oldbfe{strategic game} (in short, a game) for $n$ players ($n >
1$) we mean a sequence
\[
(S_1, \LL, S_n, p_1, \LL, p_n),
\] 
where for each $i \in [1..n]$

\begin{itemize}
\item $S_i$ is the non-empty, finite set of \oldbfe{strategies} 
available to player $i$,

\item $p_i$ is the \oldbfe{payoff function} for the  player $i$, so
$
p_i : S_1 \times \LL \times S_n \myra \cal{R},
$
where $\cal{R}$ is the set of real numbers.
\end{itemize}

Given a sequence of sets of strategies $S_1, \LL, S_n$ and $s \in S_1
\times \LL \times S_n$ we denote the $i$th element of $s$ by $s_i$,
denote $(s_1, \LL, s_{i-1}, s_{i+1}, \LL, s_n)$ by $s_{-i}$ and
similarly with $S_{-i}$, and write $(s'_i, s_{-i})$ for $(s_1, \LL,
s_{i-1}, s'_i, s_{i+1}, \LL, s_n)$, where we assume that $s'_i \in
S_i$.  We denote the strategies of player $i$ by $s_i$, possibly with
some superscripts.

Given a finite non-empty set $A$ we denote by
$\Delta A$ the set of probability distributions over $A$ and call
any element of $\Delta S_i$ a \oldbfe{mixed strategy} of player $i$.
The payoff functions are extended in the standard way to mixed strategies.

We say that $G := (S_1, \LL, S_n)$ is a
\oldbfe{restriction} of a game $H := (T_1, \LL, T_n,$ $p_1, \LL,p_n)$ if
each $S_i$ is a (possibly empty) subset of $T_i$.
We identify the restriction $(T_1, \LL, T_n)$ with $H$.
A \oldbfe{subgame} of $H$ is a restriction $(S_1, \LL, S_n)$ with all $S_i$ non-empty.

To analyze various ways of iterated elimination of strategies from an
initial game $H$ we view such procedures as operators on the set of
subgames of $H$.  A minor complication is that this set together with
the componentwise inclusion on the players' strategy sets 
does not form a lattice. Consequently, we
extend these operators to the set of all restrictions of $H$, which
together with the componentwise set inclusion does form a lattice.

Given a restriction $G := (S_1, \LL, S_n)$ of $H = (T_1, \LL, T_n,
p_1, \LL, p_n)$ and two strategies $s_i \in T_i$ and $m_i \in \Delta
T_i$ we write $m_i \succ_{G} s_i$ as an abbreviation for 
\[
\fa s_{-i} \in S_{-i} \: p_{i}(m_i, s_{-i}) > p_{i}(s_i, s_{-i})
\]
and $m_i \succ^w_{G} s_i$ as an abbreviation for
\[
\fa s_{-i} \in S_{-i} \: p_{i}(m_i, s_{-i}) \geq p_{i}(s_i, s_{-i}) \A \te s_{-i} \in S_{-i} \: p_{i}(m_i, s_{-i}) > p_{i}(s_i, s_{-i}).
\]
In the first case we say that $s_i$ \oldbfe{is strictly dominated} \oldbfe{on} $G$
\oldbfe{by} $m_i$ and in the second one that
$s_i$ \oldbfe{is weakly dominated} \oldbfe{on} $G$
\oldbfe{by} $m_i$. In particular $m_i$ can be a pure strategy, i.e.~an element of $T_i$.

Given an operator $T$ on a finite lattice $(D, \sse)$
we denote by $T^{k}$ the $k$-fold iteration of $T$, where $T^0 = D$ 
(so the iterations start 'at the top') and 
let $T^{\omega} := \cap_{k \geq 0} T^k$.
We call $T$ \oldbfe{monotonic} if for all $G, G'$
\[
\mbox{$G \sse G'$ implies $T(G) \sse T(G')$}.
\]

When comparing two ways of eliminating strategies from a strategic
game, represented by the operators $T$ and $U$ on the lattice of all
restrictions of $H$, we would like to deduce $T^{\omega} \sse
U^{\omega}$ from the fact that for all $G$, $T(G) \sse U(G)$.
Unfortunately, in general this implication does not hold; a revealing
example is provided in Section \ref{sec:weak}.  What does hold is the
following simple lemma that relates to steps (iii) and (iv) of the
generic procedure from the Introduction and reveals the importance of
the monotonicity.

\begin{lemma} \label{lem:inc}
Consider two operators $T$ and $U$ on a finite lattice $(D, \sse)$, such that
\begin{itemize}
\item for all $G$, $T(G) \sse U(G)$,

\item at least one of $T$ and $U$ is monotonic.

\end{itemize}
Then $T^{\omega} \sse U^{\omega}$.
\end{lemma}
\Proof
We prove by induction 
that for all $k \geq 0$ we have $T^{k} \sse U^{k}$.
The claim holds for $k = 0$.
Suppose it holds for some
$k$. Then by the assumptions and the induction
hypothesis we have the following string of inclusions and equalities:

\begin{itemize}
\item if $T$ is monotonic: $T^{k + 1} =   T(T^{k}) \sse T(U^{k}) \sse U(U^{k}) = U^{k + 1}$,

\item if $U$ is monotonic: $T^{k + 1} =   T(T^{k}) \sse U(T^{k}) \sse U(U^{k}) = U^{k + 1}$.
\HB
\end{itemize}

\section{Strict dominance}
\label{sec:strict}

From now we fix an initial game $H = (T_1, \LL, T_n, p_1, \LL, p_n)$. 
Given a restriction $G := (S_1, \LL, S_n)$ of $H$ we denote $S_i$ by $G_i$.
In particular we denote $T_i$ by $H_i$.

First we focus on two operators on the restrictions of $H$:
\[
\LS(G) := G',
\]
where for all $i \in [1..n]$
\[
G'_i := \{ s_i \in G_i \mid \neg \te s'_i \in G_i \: s'_i \succ_{G} s_i\},
\]
and $\MLS$ defined analogously but with 
$G'_i := \{ s_i \in G_i \mid \neg \te m_i \in \Delta G_i \: m_i \succ_{G} s_i\}$.

Starting with \cite{LR57}, the iterated elimination of strictly
dominated strategies is customarily defined as the outcome of the
iteration of the $\MLS$ operator starting with the initial game $H$.

To reason about the above two operators we introduce two
related operators defined by ('$G$' stands for `global'):
\[
\GS(G) := G'
\]
where for all $i \in [1..n]$
\[
G'_i := \{ s_i \in G_i \mid \neg \te s'_i \in H_i \: s'_i \succ_{G} s_i\}.
\]
and $\MGS$ defined analogously but with $G'_i := \{ s_i \in G_i \mid \neg \te m_i \in \Delta H_i \: m_i \succ_{G} s_i\}$.

So in the $\LS$ and $\MLS$ operators we limit our attention to strict
dominance by a pure/mixed strategy in the \emph{current} game, $G$,
while in the other two operators we consider strict dominance by a
pure/mixed strategy in the \emph{initial} game, $H$.

The difference is crucial because 
the operators $\LS$ and $\MLS$ are not monotonic, 
while $\GS$ and $\MGS$ are.
To see the former just take the following game $H$:
\begin{center}
\begin{game}{2}{1}
    &    $X$  \\
$A$ &    $1,0$ \\
$B$ &    $0,0$
\end{game}
\end{center}
Note that $\LS(H)\!=\!\MLS(H)\!=\!(\{A\}, \{X\})$ and
$\LS(\{B\}, \{X\})\!=\!\MLS(\{B\}, \{X\})$ $= (\{B\}, \{X\})$.
So $(\{B\}, \{X\}) \sse H$, while neither $\LS(\{B\}, \{X\}) \sse \LS(H)$
nor 
$\MLS(\{B\}, \{X\}) \sse \MLS(H)$.

Monotonicity of $\GS$ and $\MGS$ follows directly from their
definitions.  The $\MGS$ operator is studied in \cite{BFK06} (it is
their operator $\Phi$) and in \cite{Apt07}.  For the case of strict
dominance by a pure strategy it was introduced for arbitrary games in
\cite[ pages 1264-1265]{MR90}, studied for compact games with
continuous payoffs in \cite[ Section 5.1]{Rit02} and considered for
arbitrary games in the presence of transfinite iterations in
\cite{CLL05} and \cite{Apt07}.

The fact that the $\MGS$ operator is monotonic has some mathematical
advantages.  For example, by virtue of a general result established in
\cite{Apt07}, it is automatically order independent. Moreover, thanks to
monotonicity, as argued in \cite{Apt07a}, it can be used in the
epistemic framework of game theory based on possibility
correspondences as `a stand alone' concept of rationality.

The following result, the second part of which was proved (as
Proposition 2.2 (ii)) in \cite{BFK06}, relates the original operators
to their global versions and corresponds to step (ii) of the generic
procedure from the Introduction.

\begin{lemma} \label{thm:finite1}
$\GS^{\omega} = \LS^{\omega}$ and $\MGS^{\omega} = \MLS^{\omega}$.
\end{lemma}
\Proof 
See the appendix.
\HB
\VV

This allows us to establish the expected inclusion.
\begin{theorem} \label{thm:m}
$\MLS^{\omega} \sse \LS^{\omega}$.
\end{theorem}
\Proof
This is a consequence of the mentioned generic
procedure, since step (iii) holds:
for all restrictions $G$, $\MGS(G) \sse \GS(G)$, and step (iv) holds:
both $\MGS$ and $\GS$ are monotonic.
\HB

\section{Weak dominance}
\label{sec:weak}

Next we compare the $\LS$ and $\MLS$ operators with their weak dominance counterparts, $\LW$ and $\MLW$,
defined in the same way, but using the $\succ_G^w$ relation instead of $\succ_G$.
In the literature, starting with \cite{LR57}, 
the iterated elimination of weakly dominated strategies is customarily defined
as the outcome of the iteration of the $\MLW$ operator starting with the initial game $H$.

To reason about these two operators we use the corresponding `global'
versions, $\GW$ and $\MGW$, defined in the same way as for strict
dominance. We have following counterpart of Lemma \ref{thm:finite1},
the second part of which was proved (as Lemma F.1) in \cite{BFK07}.

\begin{lemma} \label{thm:finite2}
$\GW^{\omega} = \LW^{\omega}$ and $\MGW^{\omega} = \MLW^{\omega}$.
\end{lemma}
\Proof 
See the appendix.
\HB
\VV

This allows us to establish the following result.

\begin{theorem} \label{thm:finite3}
$\LW^{\omega} \sse \LS^{\omega}$ and $\MLW^{\omega} \sse \MLS^{\omega}$.
\end{theorem}
\Proof 
Again, this is a consequence of the generic
procedure from the Introduction. Indeed, step (iii) holds:
for all restrictions $G$, $\GW(G) \sse \GS(G)$ and $\MGW(G) \sse \MGS(G)$,
and step (iv) holds: both $\GS$ and $\MGS$ are monotonic.
%

\HB
\VV

However, surprisingly, the inclusion $\MLW^{\omega} \sse \LW^{\omega}$ does not hold.

 \begin{example} \label{exa:weak}
 Consider the following game $H$:
  
 \begin{center}
 \begin{game}{4}{3}
       & $X$   & $Y$   & $Z$ \\
 $A$   &$2,1$  & $0,1$ & $1,0$ \\
 $B$   &$0,1$  & $2,1$ & $1,0$ \\
 $C$   &$1,1$  & $1,0$ & $0,0$ \\
 $D$   &$1,0$  & $0,1$ & $0,0$
 \end{game}
 \end{center}
Applying to it the $\MLW$ operator we get
 \begin{center}
 \begin{game}{2}{2}
       & $X$    & $Y$   \\
 $A$   & $2,1$  & $0,1$ \\
 $B$   & $0,1$  & $2,1$
 \end{game}
 \end{center}
Another application of $\MLW$ yields no change.  In contrast, after
three iterations of the $\LW$ operator to the initial game we reach
\begin{center}
 \begin{game}{1}{1}
       & $X$  \\
 $A$   &$2,1$ 
 \end{game}
 \end{center}
\HB
\end{example}

Since the conclusion of Lemma \ref{lem:inc} does not apply here, while Lemma \ref{thm:finite2}
holds, we conclude that none of the operators
$\LW, \MLW, \GW$ and $\MGW$ is monotonic.

\section*{Acknowledgement}
We acknowledge helpful email discussions with Adam Brandenburger.

\bibliography{/ufs/apt/bib/e,/ufs/apt/bib/apt}
\bibliographystyle{handbk}

\section*{Appendix}

\textbf{Proof of Lemma \ref{thm:finite1}.}

This result is a corollary to Theorem 5 of \cite{Apt07}. We provide here a direct proof.
Note first that for all restrictions $G$ we have $\GS(G) \sse \LS(G)$ so
by Lemma \ref{lem:inc} we have $\GS^{\omega} \sse \LS^{\omega}$.

To establish the first equality we prove by induction that
for all $k \geq 0$
\[
\LS^{k} \sse \GS^{k}.
\]
We only need to establish the induction step.  
Take $s_i \in \LS_i^{k+1}$.  
By definition $s_i \in \LS_i^{k}$ and $\neg \te s'_i \in
\LS^{k}_i \: s'_i \succ_{\LS^{k}} s_i$. By the induction hypothesis
$s_i \in \GS_i^{k}$.
Let
\[
A := \{s'_i \in H_i \mid s'_i \succ_{\LS^{k}} s_i\}.
\]
Suppose by contradiction that $A \neq \ES$.  
Choose a maximal element $s^{*}_i \in A$ w.r.t.~the
$\succ_{\LS^{k}}$ ordering on $A$.
Then for all $l \in [0..k]$, $s^{*}_i \in \LS_{i}^l$, 
so in particular $s^{*}_i \in \LS_{i}^k$,
which contradicts the assumption about $s_i$. So $A = \ES$,
which implies by the induction hypothesis that $\neg \te s'_i \in
H_i \: s'_i \succ_{\GS^{k}} s_i$. So 
$s_i \in \GS_i^{k+1}$.
\HB
\VV

\NI
\textbf{Proof of Lemma \ref{thm:finite2}.}

We prove by induction that for all $k \geq 0$
\[
\LW^{k} = \GW^{k}.
\]
Again, we only need to establish the induction step.  
We prove first that $\LW^{k+1} \sse \GW^{k+1}$.
Take $s_i \in \LW^{k+1}_{i}$.
By the induction hypothesis $\LW^{k} = \GW^{k}$, so $s_i \in \GW^{k}_{i}$.
Let
\[
A := \{s'_i \in H_i \mid s'_i \succ_{\LW^{k}} s_i\}.
\]

Suppose by contradiction that $A \neq \ES$. 
Define the function $\emph{index}: A \myra [0..k]$ by
\[
\emph{index}(s'_i) := \left\{ 
\begin{tabular}{ll}
$k$ &  \mbox{if $s'_i \in \LW^{k}_{i}$} \\[2mm]
$j$ &  \mbox{if $s'_i \in \LW^{j}_{i} \setminus \LW^{j+1}_{i}$} 
\end{tabular}
\right .  \\
\]

Let $B$ be the set of elements of $A$ with the maximal index, i.e.,
\[
B = \textrm{arg} \max_{s'_i \in A} \emph{index}(s'_i).
\]
Let now $j_0$ be the \emph{index} of the elements in $B$ and $s^{*}_i$ a
maximal element of $B$ w.r.t.~the $\succ^{w}_{\LW^{j_0}}$ ordering on $B$.

Note that $j_0 = k$.  Indeed, otherwise $s^{*}_i \not\in
\LW^{j_{0}+1}_i$. Then for some $s''_i \in \LW^{j_{0}}_i$ we have
$s''_i \succ^{w}_{\LW^{j_0}} s^{*}_i$. But $s^{*}_i \succ_{\LW^{k}} s_i$ and $\LW^{k} \sse \LW^{j_0}$, so
$s''_i \succ^{w}_{\LW^{k}} s_i$.
Hence $s''_i \in B$, which 
contradicts the choice of $s^{*}_i$.

So $j_0 = k$, which means that $s^{*}_i \in {\LW_i^{k}}$ and 
$s^{*}_i \succ^{w}_{\LW^{k}} s_i$. But this contradicts the fact that 
$s_i \in \LW^{k+1}_{i}$. So $A = \ES$, which implies by the induction hypothesis
that $\neg \te s'_i \in H_i \: s'_i \succ^{w}_{\GW^{k}} s_i$. So $s_i \in \GW^{k+1}_{i}$.

Next we prove that $\GW^{k+1} \sse \LW^{k+1}$.
Take $s_i \in \GW^{k+1}_i$. Then $s_i \in \GW^{k}_i = \LW^{k}_i$. Also
$\neg \te s'_i \in H_i \: s'_i \succ^{w}_{\GW^{k}} s_i$, so
$\neg \te s'_i \in \LW^{k}_i \: s'_i \succ^{w}_{\GW^{k}} s_i$, 
so by the induction hypothesis
$\neg \te s'_i \in \LW^{k}_i \: s'_i \succ^{w}_{\LW^{k}} s_i$, which means
that $s_i \in \LW^{k+1}_i$.

\HB

\end{document}